# Looking for a compact semi empirical equation of state for hard spheres: possibility of a glassy transition and random loose packing

by Richard BONNEVILLE

Centre National d'Etudes Spatiales (CNES), 2 place Maurice Quentin, 75001 Paris, France

phone: +33 6 87 60 78 41, mailto:richard.bonneville@cnes.fr

**Abstract**

The equation of state of a hard sphere fluid at high density should exhibit a simple pole at the random close packing limit. Here we show that trying to obtain a compact semi-empirical equation of state simultaneously compatible with that asymptotic behaviour and with the known virial coefficients raises an analytical difficulty which can be solved if a glassy transition occurs in the disordered metastable phase at a density intermediate between the freezing point and the random close packing limit, numerically close to the melting point. The estimated value for the transition point, which is identified with random loose packing, is in good agreement with earlier estimations.

## Introduction

Numerical simulations [1,2] show that at high density in the disordered metastable phase the equation of state of an hard sphere system should evidence a simple pole for a value $\xi_0$ of the reduced density parameter $\xi = \frac{\mathrm{Nv}}{\mathrm{V}}$ which is identified with the random close packing limit ( $\mathrm{V}$ : volume, $\mathrm{N}$ : number of particles, $\mathrm{v}$ molecular volume), i.e. the asymptotic behaviour of the equation of state is

$$\lim_{\xi \to \xi_0} \frac{\mathrm{PV}}{\mathrm{Nk_B T}} = \frac{\mathrm{A}}{1 - \xi / \xi_0} \qquad (1)$$

( $\mathrm{P}$ : pressure, $\mathrm{k_B}$ : Boltzman constant, $\mathrm{T}$ : temperature).

Speedy [3] has proposed the value $\mathrm{A} \cong 2.765$ .

The precise definition of random close packing has been debated [4,5]; let us simply admit it is the "maximally random jammed state" [6]. As determined by experiments [7,8] and numerical simulations [9,10] $\xi_0$ is close to 0.64. Finney [11] has proposed the more precise value 0.637, a value close to $2/\pi$ but the identification $\xi_0 \equiv 2/\pi$ , although highly tempting, has not been proven to date [12,13].

At lower densities, a very popular equation of state of the fluid phase is the Carnahan & Starling expression [14] which is fairly accurate along the fluid branch below the freezing point ( $\xi_f \cong 0.494$ ) and even beyond up to a density close to that of the melting point ( $\xi_m \cong 0.545$ ) above which it goes totally wrong. In its domain of validity it provides an estimation of the virial coefficients ( $\mathrm{B}_p \cong (p-1)(p+2)$ for $p \geq 2$ ) in excellent agreement with the available numerical data.

## A compact semi empirical equation of state

With those beginnings, is it possible to build an equation of state valid in the whole fluid phase from 0 to $\xi_0$ and compatible with the asymptotic behaviour equ.(1)? In order to account for the pole in $\xi = \xi_0$ that hypothetical equation of state should have the following form:

$$\frac{\mathrm{PV}}{\mathrm{Nk_B T}} = 1 + 4\xi + 10\xi^2 \frac{\mathrm{f_1}(\xi)}{1 - \xi/\xi_0} \qquad (2)$$

We develop $\mathrm{f_1}(\xi)$ as a series of the density, i.e.

$$\frac{\mathrm{PV}}{\mathrm{Nk_B T}} = 1 + 4\xi + 10\xi^2 \left( \frac{1 + \mathrm{a_1}\xi + \mathrm{a_2}\xi^2 + \mathrm{a_3}\xi^3 + \mathrm{a_4}\xi^4 + \ldots}{1 - \xi/\xi_0} \right). \qquad (3)$$

The connection between the $\mathrm{a_q}$ and the virial coefficients $\mathrm{B_p}$ is

$$\mathrm{a_q} = \mathrm{b_p} - (\xi_0)^{-1} \mathrm{b_{p\text{-}1}} \qquad (4)$$

with $q = p - 3$ and $b_p = B_p / 10$.

For the virial coefficients we take the numerical results by Clisby & Mc Coy for $p = 1$ to $p = 10$ [15] and those published by Wheatley for $p = 11$ and $p = 12$ [16].

With the assumption $\xi_0 \equiv 2/\pi$ we obtain the $a_q$ of Table 1, 3rd column. It is easily checked that those results are only weakly dependent on the precise value chosen for $\xi_0$ in the vicinity of 0.64.

It could be imagined that in addition to the pole in $\xi = \xi_0$ the equation of state could exhibit another single pole at the face-centred cubic density $\xi_{fcc}$; that assumption has been checked and is not constructive. On the contrary by examining the successive $a_q$, it seems that their values get closer to the sequence of the natural integers as their index increases; the equation of state could thus be written as

$$\frac{PV}{Nk_BT} \cong 1 + 4\xi + 10\xi^2 \left( \frac{1}{1 - \xi/\xi_0} \right) \left( 1 + a'_1\xi + a'_2\xi^2 + a'_3\xi^3 + a'_4\xi^4 + \ldots - \xi^5 \left( 1 + 2\xi + 3\xi^2 + 4\xi^3 + 5\xi^4 + \ldots \right) \right) \tag{5}$$

We assume that the last bracket can be re-summed so that

$$\frac{PV}{Nk_BT} \cong 1 + 4\xi + 10\xi^2 \left( \frac{1}{1 - \xi/\xi_0} \right) \left( 1 + a'_1\xi + a'_2\xi^2 + a'_3\xi^3 + a'_4\xi^4 + \ldots - \frac{\xi^5}{(1-\xi)^2} \right) \tag{6}$$

That equation of state exhibits a single pole for $\xi = \xi_0$ and a double pole for $\xi = 1$; we remember that the Carnahan & Starling equation of state exhibits a triple pole for $\xi = 1$. Equ.(6) can then be expressed as

$$\frac{PV}{Nk_BT} \cong 1 + 4\xi + 10 \left( \frac{1}{1 - \xi/\xi_0} \right) \left( \frac{\xi}{1-\xi} \right)^2 \left( 1 + a''_1\xi + a''_2\xi^2 + a''_3\xi^3 + a''_4\xi^4 + a''_5\xi^5 + \ldots \right) \tag{7}$$

The computed $a''_q$ for $\xi_0 \equiv 2/\pi$ are displayed in Table 1, 4th column, and in Fig.1.

We have $a''_1 \cong -1.7343$ and $a''_2 \cong +0.4063$. The terms $a''_3\xi^3$, $a''_4\xi^4$, $a''_5\xi^5$, which are one order of magnitude smaller than the preceding ones (see 5th column of Table 1), can be accounted for by the expression

$$-0.0617\xi^3 \left( 1 + 0.8\left(\xi/\xi_0\right) + 0.9\left(\xi/\xi_0\right)^2 \right).$$

The following terms are still one order of magnitude smaller and will be neglected. That leads to the approximate equation of state

$$\frac{PV}{Nk_BT} \cong 1 + 4\xi + 10 \left( \frac{1}{1 - \xi/\xi_0} \right) \left( \frac{\xi}{1-\xi} \right)^2 \left( 1 - 1.7343\xi + 0.4063\xi^2 - 0.0617\xi^3 \left( 1 + 0.8\left(\xi/\xi_0\right) + 0.9\left(\xi/\xi_0\right)^2 \right) \right) \tag{8}$$

## Discussion

The virial coefficients derived from equ.(8) are by construction in excellent agreement with the numerical data. From equ.(8) in the high density limit the residue of the pole in $\xi_0$ is

$$A \cong 10\left[\left(\frac{\xi}{1-\xi}\right)^2\left(1-1.7343\xi+0.4063\xi^2-0.0617\xi^3\left(1+0.8\left(\xi/\xi_0\right)+0.9\left(\xi/\xi_0\right)^2\right)\right)\right]_{\xi=\xi_0} \qquad (9)$$

We find $A \approx 0.5$ whereas the expected value is $A \approx 2.8$. That discrepancy can be explained as follows. If we follow the curve of the liquid phase as given by the equation of state equ.(8) when density increases beyond the freezing point $\xi_f$, it meets the curve associated with the asymptotic equation of state equ.(1) with $A \cong 2.765$ at a density $\xi_g \approx 0.545 \pm 0.005$. That density is also that of the melting point and where the Carnahan & Starling equation of state starts failing. Then beyond $\xi_g$ up to $\xi_0$ the system is described by the asymptotic equation of state equ.(1) with $A \cong 2.765$ and no longer by equ.(8). In $\xi = \xi_g$ the pressure $P(\xi)$ is continuous but the compressibility $\partial P / \partial \xi$ is not and that is the signature of the glassy transition pointed out in ref.[1,2,3]. That transition is actually rather fuzzy and the transition point $\xi = \xi_g$ can be identified with the random loose packing density [17,18,19] defined as being the "loosest possible random packing that is mechanically stable that one can achieve by pouring grains" |20]; to support this view the value $\xi_g \approx 0.545 \pm 0.005$ of the transition point is in good agreement with earlier estimations of the random loose packing density [20,21 ].

The behaviour of the disordered phase is illustrated by Fig. 2, which shows a comparison around the glassy transition of the Carnahan & Starling's equation of state, Speedy's equation of state, and our equation of state equ.(8).

## Conclusion

Our attempt to obtain a compact semi empirical equation of state for a disordered hard sphere system suggests the evidence of a glassy transition at a density intermediate between the freezing point and the random close packing limit, numerically close to the melting point; the transition point can be interpreted as random loose packing.

**Table caption:**

Table 1: The known virial coefficients $B_p$ (ref. 15 &16) and the coefficients $a_q$ , $a''_q$ defined in this paper.

**Figure captions:**

Fig.1: The coefficients $a''_q$ as function of their index.

Fig.2: Zooming of the 3 equations of state around the glassy transition; lower (dashed) line: Carnahan & Starling’s equation of state, upper (dotted/dashed) line: Speedy’s equation of state, continuous line: our equation of state equ.(8).

Table 1

| **q** | **$B_{q+3}$** (ref. 14,15) | **$a_q$** | **$a''_q$** | **$a''_q \xi_0^q$** |
|---|---|---|---|---|
| 0 | 10 | 1 | 1 | 1.0000 |
| 1 | 18.3648 | 0.2657 | -1.7343 | -1.1041 |
| 2 | 28.2245 | -0.0623 | 0.4063 | 0.1647 |
| 3 | 39.8151 | -0.4520 | -0.0617 | -0.0159 |
| 4 | 53.3444 | -0.9197 | -0.0780 | -0.0128 |
| 5 | 68.5375 | -1.5256 | -0.1381 | -0.0144 |
| 6 | 85.8128 | -2.1846 | -0.0531 | -0.0035 |
| 7 | 105.7751 | -2.9019 | -0.0584 | -0.0025 |
| 8 | 127.9263 | -3.8225 | -0.2032 | -0.0055 |
| 9 | 150.9949 | -4.9951 | -0.2521 | -0.0043 |

Figure 1

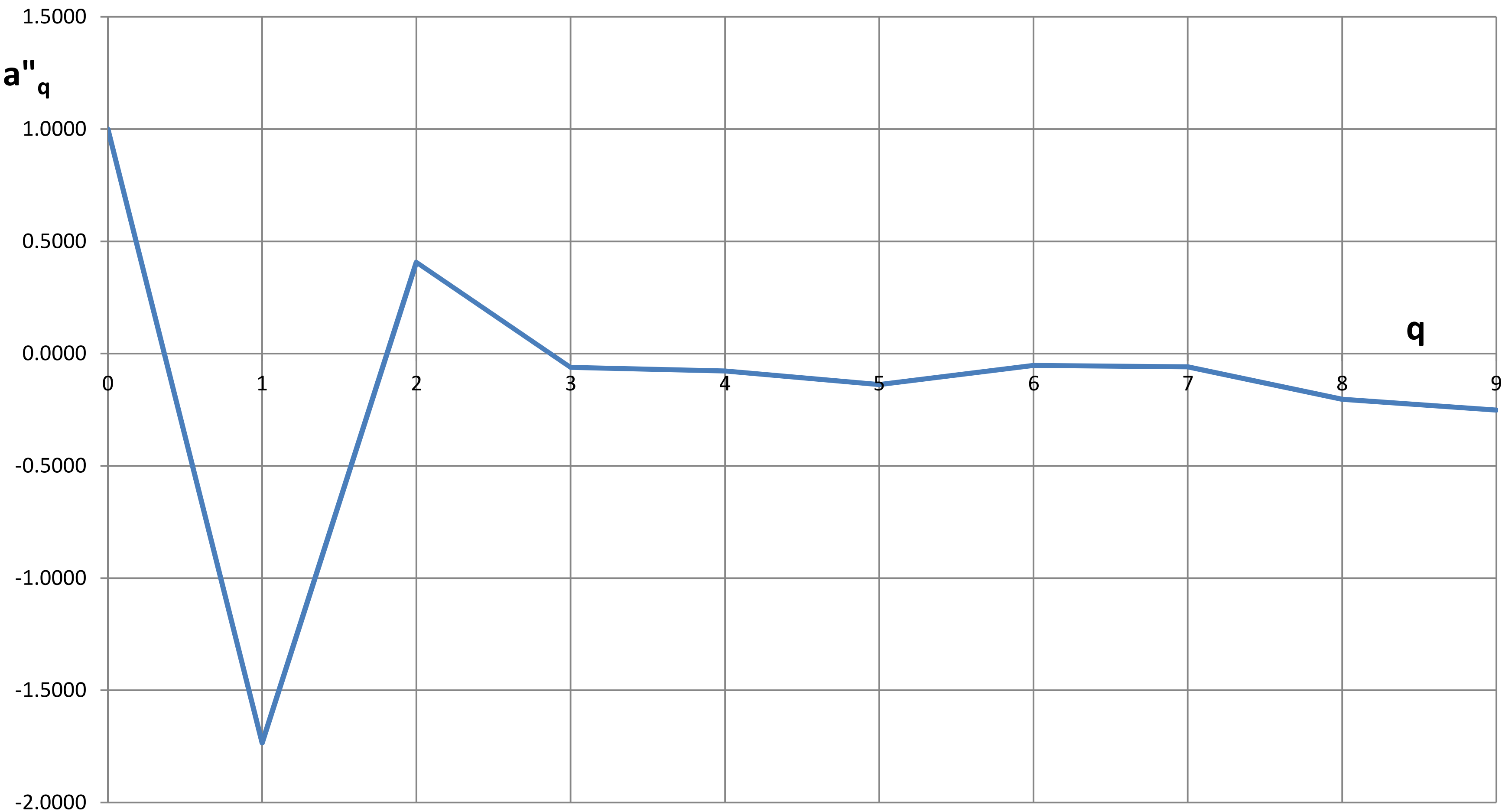

Figure 2

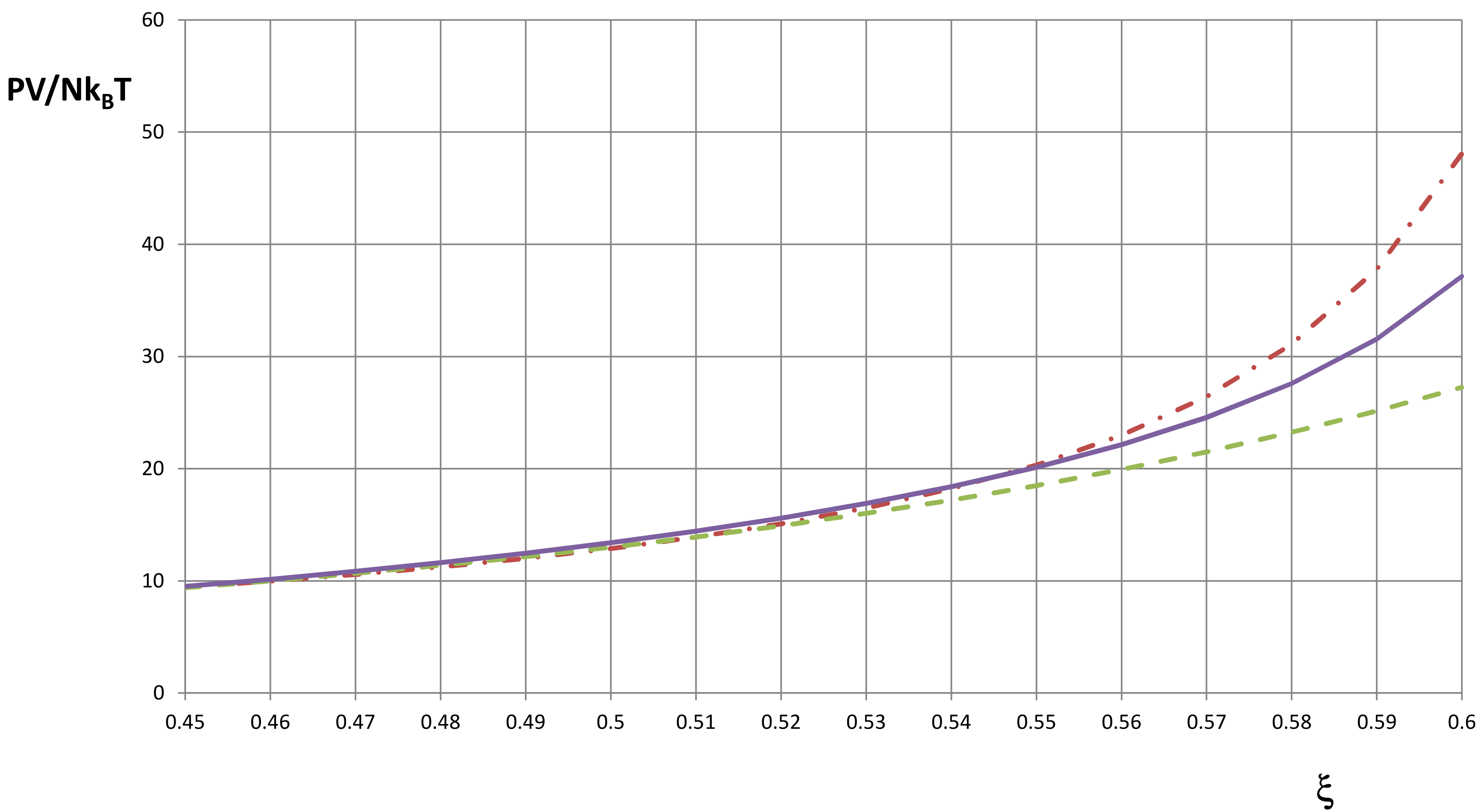